\documentstyle[12pt]{article}
\advance\textheight by 60pt
\advance\voffset by -40pt
\advance\textwidth by 50pt
\advance\oddsidemargin by -25pt
\advance\evensidemargin by -25pt

\def\single_space{\baselineskip 12pt plus 1pt minus 1pt}
\def\one_and_a_half_space{\baselineskip 19pt plus 1pt minus 1pt}
\def\double_spacesp{\baselineskip 25pt plus 2pt minus 2pt}

\newcommand{\geqnew}{\stackrel{>}{\!\ _{\sim}}}
\newcommand{\leqnew}{\stackrel{<}{\!\ _{\sim}}}

\begin{document}
\begin{titlepage}
\begin{flushright}
{\bf
PSU/TH/197\\
April 1998
}
\end{flushright}
\vskip 1.5cm
{\Large
{\bf
\begin{center}
Eliminating the low-mass axigluon window
\end{center}
}}
\vskip 1.0cm
\begin{center}
M.~A.~Doncheski \\
Department of Physics \\
The Pennsylvania State University\\
Mont Alto, PA 17237  USA \\
\vskip 0.1cm
and \\
\vskip 0.1cm
R.~W.~Robinett \\
Department of Physics\\
The Pennsylvania State University\\
University Park, PA 16802 USA\\
\end{center}
\vskip 1.0cm
\begin{abstract}
 
Using recent collider data, especially on the hadronic width the $Z^0$, 
we exclude axigluons in the currently allowed low-mass window, namely 
axigluons in the mass range $50\,GeV < M_A < 120\,GeV$.  Combined with 
hadron collider data from di-jet production, axigluons with masses below 
roughly $1\,TeV$ are now completely excluded.

\end{abstract}

\end{titlepage}
\double_spacesp

Models which extend the standard color gauge group to $SU_3(L) \times SU_3(R)$
at high energies, so-called chiral color theories, include a wide variety of
new particles beyond the standard model with the exact 
spectrum depending on the
details of the theory.   All such models, however, necessarily predict 
the existence of a massive, color-octet gauge boson, the axigluon $A$, which 
couples to quarks with an axial vector structure and the same strong 
interaction coupling
 strength as QCD. The original models \cite{glashow} explicitly connected the 
scale of the breakdown of chiral color to ordinary QCD with the electroweak 
scale and so even more specifically predicted that axigluons should have 
masses no larger than $M_A \leqnew 300\,GeV$.  Early bounds from $\Upsilon$ 
decays \cite{cuypers_1} quickly found the limit $M_A \geqnew 25\,GeV$, 
while analyses 
of limits arising 
from axigluon contributions  to the hadronic cross-section in 
$e^+e^-$ reactions (the $R$ ratio) \cite{cuypers_2} gave the limit 
$M_A > 50\,GeV$ (at $95\%$ confidence level) using then current PEP/PETRA data.
Early suggestions \cite{glashow} that axigluons might be visible as an
enhancement in the di-jet cross-section at hadron colliders were first used
by Bagger, Schmidt, and King \cite{bagger} and then by 
the UA1 collaboration \cite{ua1}
to exclude axigluons in the mass range $125 \, GeV < M_A < 310 \,GeV$.  More
recent searches for structure in the jet-jet invariant mass at the TEVATRON
have led to dramatically enhanced limits, especially for heavy axigluon 
masses, with CDF data \cite{cdf} now excluding axigluons in the range 
$120\,GeV < M_A < 980\,GeV$.  These are the limits which appear in the
Particle Data Group discussion of bounds on the axigluon mass \cite{pdg} 
and we note
that there is still a window of allowed masses in the range
$50\,GeV < M_A < 120\,GeV$ which has not yet been excluded. This 
allowed window constitutes a large fraction of the range in masses
(namely, up to $\sim 300\,GeV$) predicted by the original models which
motivated the searches for axigluons and improved limits 
 in this region would be useful in testing chiral color theories. 
In this note,
we will use several rather different types of recent collider data to
exclude axigluons in this mass region for the first time.
  Combined with the di-jet
limits, this will imply that axigluons with masses in the entire
range below roughly $1\,TeV$ will be definitively excluded.
  To the extent that chiral
color models are constructed with the scale of color breaking directly tied
to the electroweak scale, all such theories are also excluded.

The present authors have recently considered the effects of axigluons
on the dominant ($90\%$ of the production cross-section) 
$q\overline{q} \rightarrow t\overline{t}$ subprocess
contributing to the top quark production cross-section \cite{top_quark} 
and noted  that the 
inclusion of axigluons with masses 
in the low-mass window more than doubles  the 
tree-level cross-section.  Even with the good agreement of the NLO
QCD predictions with the CDF and D0 data, the inclusion of axigluons
in the low-mass window is not yet definitively excluded due to the rather
large experimental errors, but is definitely disfavored at the 
$(1\!-\!1.5)\sigma$ level.   Given the large NLO QCD corrections to the
tree-level $q\overline{q}$ process, one might well imagine that a complete
NLO analysis, including the effects of low-mass axigluons, would make
an unacceptably large contribution to the $t\overline{t}$ cross-section.
It was pointed out some time ago, however, that relatively light axigluons
can disturb the perturbative calculability of tree-level partial wave
amplitudes \cite{partial_wave} 
for processes involving heavy quarks, so that top quarks  coupled
to sufficiently light axigluons would be strongly interacting.  Extending
work by Chanowitz, Furman, and Hinchliffe \cite{ultra_heavy} on the
interactions of ultra-heavy fermions, one of the present authors
\cite{partial_wave} found that the $J=0$ tree-level partial-wave amplitude
for $Q\overline{Q} \rightarrow Q \overline{Q}$ (via $s$- and
$t$-channel axigluon exchange) would become non-perturbative (i.e.
$|a_0| > 1$) unless the axigluon mass satisfied the inequality
$M_A > \sqrt{5\alpha_s/3} M_Q$.  Using the measured value of the top
quark mass and the apparently very good agreement between the NLO 
(perturbative!) QCD
predictions for the top-quark production cross-sections and the experimentally
observed value, this implies that $M_A \geqnew  72\, GeV$, which already 
improves the older $e^+e^-$ bound, pushing the limit closer to the
$Z^0$ mass.

Other more specialized collider data might also be used to bound the axigluon
mass.  The associated production of an axigluon with a weak boson via
the subprocess $q+ \overline{q}' \rightarrow W/Z + A$ is similar to that
used for the production of the standard model Higgs boson via
$p \overline{p} \rightarrow W/Z + X^0$ with $X^0 = H^0 \rightarrow b
\overline{b}$.  Given the expected large branching ratio of the axigluon
to $b\overline{b}$ final states ($BR(A \rightarrow b\overline{b})
 = 1/5$ for $5$ active quark flavors) and the
much larger coupling of the axigluon to the initial quarks, this channel
might easily be used to extract limits on $M_A$.

The tree-level partonic cross-section for the largest subprocess, 
namely $q\overline{q}' \rightarrow W A$,  is easily found to be
\begin{equation}
\frac{d \hat{\sigma}}{d\hat{t}}
(q\overline{q}' \rightarrow W A)
= \frac{4\alpha_s}{9}
\left[\frac{G_F M_W^2}{\sqrt{2}}\right]
\frac{|V_{qq'}|^2}{\hat{u}\hat{t}\hat{s}^2}
\left[\hat{u}^2 + \hat{t}^2 + 2\hat{s}(M_W^2 + M_A^2)
- 
\frac{M_A^2 M_W^2(\hat{u}^2+\hat{t}^2)}{\hat{u}\hat{t}} \right]
\end{equation}
and at TEVATRON energies we find the following values for total
cross-sections times branching ratios (assuming $BR(A \rightarrow
b\overline{b}) = 0.2$) for various axigluon masses in the allowed
window:

\begin{equation}
\begin{array}{cccc} 
M_A         & \sigma \cdot BR  & M_A        & \sigma \cdot BR  \\ \hline
50\,GeV     &  51\,pb          &  90 \,GeV  &    26\,pb  \\
60\,GeV     &  46\,pb          &  100\,GeV  &    17\,pb  \\
70\,GeV     &  40\, pb         &  110\,GeV  &    16\,pb  \\
80\,GeV     &  28\, pb         &  120\,GeV  &    14\,pb  \\
\end{array}
\end{equation}
Bhat \cite{bhat} has recently surveyed limits for many new physics searches
at the TEVATRON and presented preliminary data for the production cross-section
times branching ratio for $p\overline{p} \rightarrow W + X^0$ with
$X^0 = H^0 \rightarrow b\overline{b}$ from CDF (which uses an $l\nu$ tag
for  the $W$ bosons).  The limits cover the mass range $75\,GeV \leqnew
M_H \leqnew 125 \,GeV$ and are of order $\sigma \cdot BR \approx 15-20\,pb$.
Given the estimates above, a complete analysis of this process for the 
axigluon analog could likely exclude axigluons up to 
$M_A \approx 80-90 \,GeV$, with $M_A > 70\,GeV$ a seemingly safe estimate
of the current bound possible. 

Given the huge statistical sample of $Z^0$ hadronic decay events at LEP,
it is perhaps most natural to extend the analyses of Ref.~\cite{cuypers_2}
using LEP data.  Following the same strategy as employed previously,
we compare the value of $\alpha_s$ extracted from low-energy experiments
(which is then run up to $M_Z$) with the value extracted from the
hadronic width of the $Z^0$ at the pole.  (The improved limits
on $M_A$ mentioned above imply that any changes to the running of
$\alpha_s$ due to axigluon effects will be small.) 
The inclusion of real and virtual
axigluons increases the hadronic decay rate (or $R$ value in $e^+e^-$
collisions) by a factor of $(1 + \alpha_s(\sqrt{s})f(\sqrt{s}/M_A)/\pi
+ {\cal O}(\alpha_s^2))$ where the function 
$f(\sqrt{s}/M_A)$ is derived in Ref.~\cite{cuypers_2}.  
The Particle Data Group perturbative QCD analysis
\cite{pdg_qcd} 
quotes a value of $\alpha_s$ derived from low-energy data (such as
deep-inelastic scattering (excluding HERA), $\tau$ decay, $\Upsilon$
width, and lattice calculations), namely, $\alpha_s^{(LE)}(M_Z)
= 0.118 \pm 0.004$.  The value of $\alpha_s$ extracted from the ratio of
hadronic to leptonic decay widths of the $Z^0$ ($\Gamma_h/\Gamma_{\mu}
= 20.788 \pm 0.0032$, which probes the same QCD corrections as the $R$
value at lower energies) is $\alpha_s^{(HE)}(M_Z) = 0.123 
\pm 0.004 \pm 0.002$ and the evaluation includes the effect of ordinary
QCD (gluonic) corrections up to ${\cal O}(\alpha_s^2)$.  Using bounds
on the possible difference between these two values, after combining
errors, we find that the contribution from axigluons, due to the 
$f(\sqrt{s}/M_A)$ term, is bounded by 
$0.042 \pm 0.05 \geq f(M_Z/M_A)$ which we take to imply roughly that 
$f(M_Z/M_A) < 0.092 \,(0.142)$ at the $1\sigma\,(2\sigma)$ or
$65\% \,(95\%)$ confidence level.  Using the expression for $f(z)$
in Ref.~\cite{cuypers_2}, we find that this corresponds to
$M_Z/M_A < 0.16 \,(0.25)$ or $M_A > 6.2M_Z \,(4M_Z)$ or
$M_A > 570\,GeV \,(365\,GeV)$ at $65\% \,(95\%)$ confidence level. 
While this is a very simple estimate, given the substantial agreement
of the various $\alpha_s$ measurements, axigluons as light as $120\,GeV$
are obviously excluded. 
Measurements of similar quantities at higher energies at LEP
(starting at $\sqrt{s} \approx 130\!-\!40\,GeV$ \cite{lep_2} and beyond the
$W^{+}W^{-}$ threshold), but with
much lower statistics, do not improve on these limits.
Nonetheless, this analysis easily excludes axigluons in the low-mass
window once and for all.  Chiral-color models and axigluons, 
if they have any relevance in nature, can only appear as new physics 
beyond the $TeV$ scale.

\begin{center}
{\Large
{\bf Acknowledgments}}
\end{center}

One of us (M.A.D) acknowledges the support of Penn State University 
through a Research Development Grant (RDG).  Both authors thank the
organizers of the Pheno - CTEQ 98 meeting, where this work
was begun, for their hospitality.

\end{document}